%% file: ISSCS2013schnurrer.tex
\documentclass[a4paper,conference]{IEEEtran}
\usepackage[latin9]{inputenc}
\usepackage{amsmath}
\usepackage{graphicx}

\makeatletter

\newcommand{\lyxdot}{.}

\usepackage{pstricks, pst-node, pst-plot, pst-circ}
\usepackage{moredefs}

\usepackage{psfrag}
\usepackage{array}

\newcommand{\HP}{\text{H\hspace{-0.25mm}P}}
\newcommand{\LP}{\text{L\hspace{-0.25mm}P}}
\newcommand{\sigf}{f}
\newcommand{\pred}{p}
\newcommand{\upd}{u}
\newcommand{\updSE}{\hat{u}}

\newcommand{\fig}{Fig.}
\newcommand{\tab}{Table}

\newcommand{\spaceBelowFig}{\vspace{-4mm}}
\newcommand{\spaceBeforeLabel}{\vspace{-2mm}}

\newcommand{\disocclusion}{dis-occlusion}

\makeatother

\begin{document}

\author{\authorblockN{Wolfgang Schnurrer, Jürgen Seiler, and André
Kaup}\\
\authorblockA{Multimedia Communications and Signal Processing\\
University of Erlangen-Nuremberg, Cauerstr. 7, 91058 Erlangen, Germany\\
Email: \{ schnurrer, seiler, kaup \}@lnt.de}}

\specialpapernotice{(Invited Paper)}

\title{Improving Block-Based Compensated Wavelet Lifting by Reconstructing
Unconnected Pixels}
\maketitle
\begin{abstract}
This paper presents a new approach for improving the visual quality
of the lowpass band of a compensated wavelet transform. A high quality
of the lowpass band is very important as it can then be used as a
downscaled version of the original signal. To adapt the transform
to the signal, compensation methods can be implemented directly into
the transform. We propose an improved inversion of the block-based
motion compensation by processing unconnected pixels by a reconstruction
method. We obtain a better subjective visual quality while furthermore
saving up to 2.6\% of bits for lossless coding.
\end{abstract}

\IEEEpeerreviewmaketitle

\section*{Introduction% no \PARstart
 }

In many video applications a scalable representation of the video
sequences is very desirable especially when originally huge data has
to be transmitted. A smaller resolution, e.g., in temporal direction,
can be used for previewing or displaying on mobile devices. Thereby,
a high quality of a downscaled representation is very important. The
wavelet transform can lead to such a scalable representation but has
the drawback of a blurry lowpass band. The lowpass band can further
contain ghosting artifacts due to motion in the video sequences.

To improve the quality of the lowpass band, compensation methods can
be incorporated directly into the transform. This technique is well
known as Motion Compensated Temporal Filtering (MCTF) for video sequences
\cite{garbasTCSVT}.

The compensation method has to be inverted in the update step of the
wavelet transform \cite{ohm1993}. When using a block-based compensation
method, blocking artifacts can occur in the lowpass band due to the
inversion procedure.

In this paper, we will improve the visual quality of the lowpass band
by avoiding annoying block artifacts caused by unconnected pixels.
Further we can reduce the filesize for lossless coding compared to
the traditional block-based motion compensated wavelet lifting.

\begin{figure}
\psfragscanon

\psfrag{ref}{$\sigf_{2t-1}$}
\psfrag{cur}{$\sigf_{2t}$}
\psfrag{pred}{$\pred_{2t}$}
\psfrag{upd}{$\upd_{2t}$}
\psfrag{hf}{$\updSE_{2t}$}
\psfrag{MC}{MC}
\psfrag{IMC}{IMC}
\psfrag{FSE}{FSE}
\psfrag{H}{$\HP_t$}
\psfrag{L}{$\LP_t$}
\psfrag{ph}{$+$} %\frac{1}{2}
\psfrag{m}{$-$}

\includegraphics[width=0.99\columnwidth]{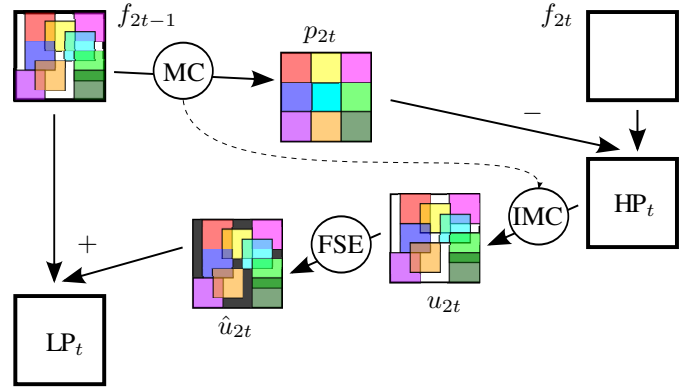}

\psfragscanoff

\spaceBeforeLabel

\caption{\label{fig:Lifting-structure}Proposed scheme: Lifting structure with
block-based compensation and Frequency Selective Extrapolation for
processing unconnected pixels}

\spaceBelowFig
\end{figure}

The following sections briefly review MCTF and block-based compensation
and introduce our proposed scheme for improving the compensated lifting.
The simulation results are discussed in Section~\ref{sec:Simulation-results}.

\section{\label{sec:Compensated-Wavelet-Lifting}Compensated Wavelet Lifting}

The lifting structure is a factorized representation of the wavelet
transform \cite{sweldens1995lifting}. \fig{}~\ref{fig:Lifting-structure}
shows a schematic of the compensated lifting structure \cite{garbasTCSVT}
of the Haar wavelet that has been extended by the Frequency Selective
Extrapolation (FSE). The highpass coefficients $\HP_{t}$ are computed
in the prediction step and the lowpass coefficients $\LP_{t}$ are
computed in the update step by using the already computed highpass
coefficients.

As illustrated in \fig{}~\ref{fig:Lifting-structure}, a motion
compensated transform is achieved by subtracting a motion compensated
(MC) predictor $\pred_{2t}$ instead of the original reference frame
$\sigf_{2t-1}$ from the current frame $\sigf_{2t}$
\begin{equation}
\HP_{t}=\sigf_{2t}-\left\lfloor \pred_{2t}\right\rfloor .\label{eq:H-Haar}
\end{equation}
To obtain an equivalent wavelet transform, the compensation has to
be inverted (IMC). So in the update step, the inverse compensated
highpass coefficients $\upd_{2t}$ are added to the reference frame.
The lowpass coefficients $\LP_{t}$ are computed to 
\begin{equation}
\LP_{t}=\sigf_{2t-1}+\left\lfloor a_{k}\cdot\upd_{2t}\right\rfloor .\label{eq:L-Haar}
\end{equation}
by using the later discussed weighting factors $a_{k}$. By further
introducing floor operators in the lifting structure, rounding errors
are avoided and the original sequence can be perfectly reconstructed
from the transform coefficients without loss \cite{calderbank1997}.
This makes the transform very feasible for high fidelity video applications
as well as for medical image data.

\section{Block-Based Compensation and its Inversion}

Compensation methods are incorporated in the wavelet transform to
obtain a high quality lowpass band without ghosting artifacts. As
in hybrid video coding, usually block-based 
\begin{figure*}
\psfragscanon

\psfrag{time}{time}
\psfrag{ft1}{$f_{2t}$}
\psfrag{ft2}{$f_{2t+1}$}
\psfrag{P}{$\mathcal{P}$}
\psfrag{U}{$\mathcal{U}$}
\psfrag{HPt}{$\HP_{t}$}
\psfrag{LPt}{$\LP_{t}$}

\includegraphics[width=0.11\textwidth]{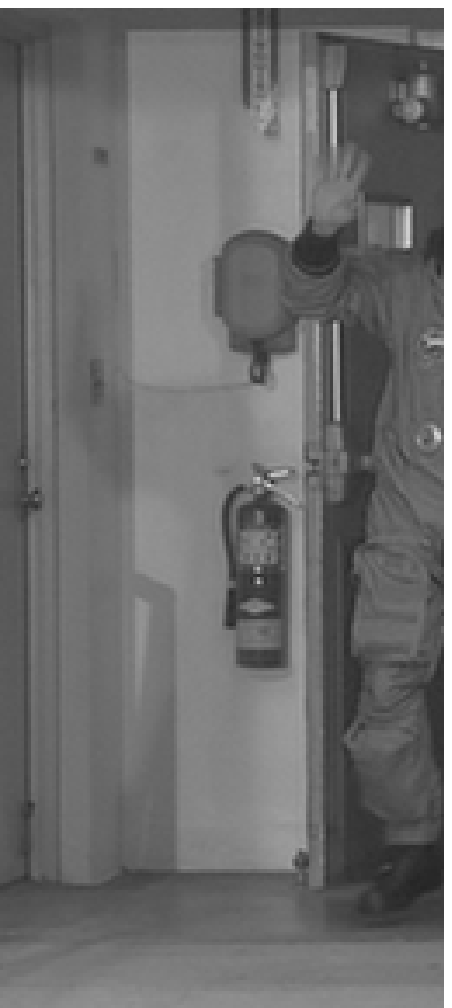}\hfill{}\includegraphics[width=0.11\textwidth]{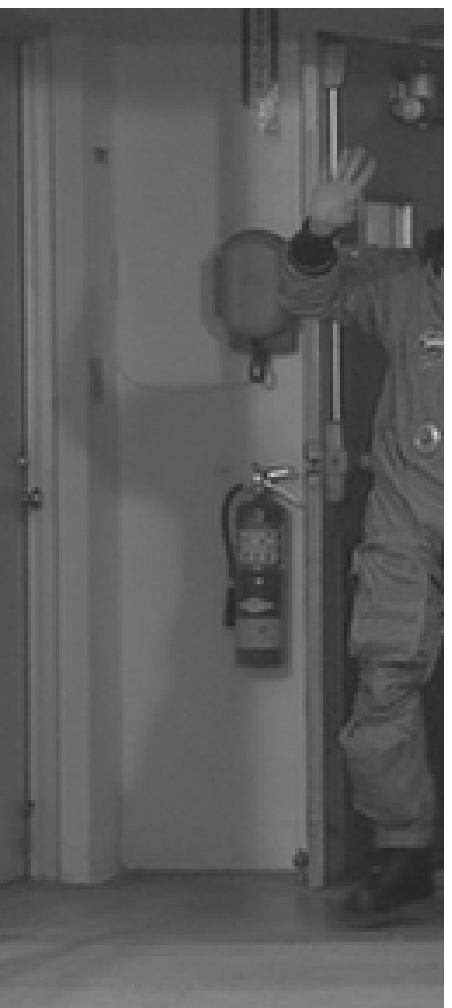}\hfill{}\includegraphics[width=0.11\textwidth]{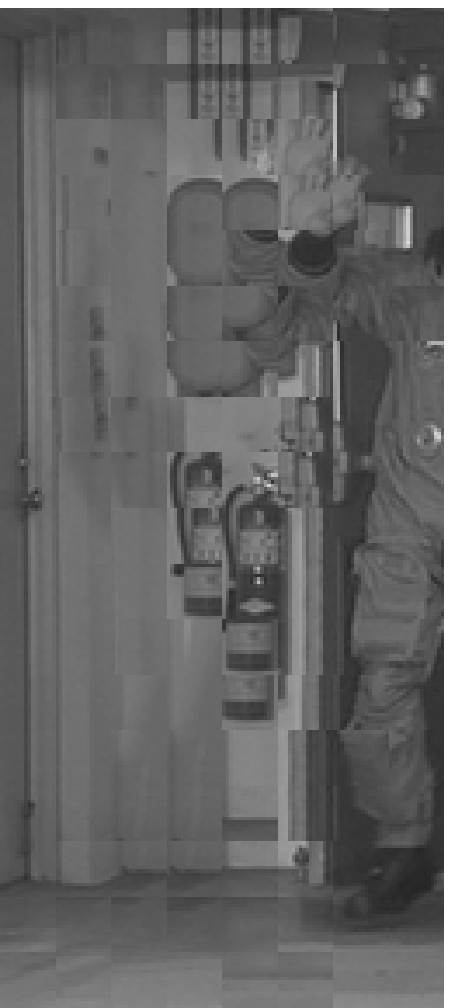}\hfill{}\includegraphics[width=0.11\textwidth]{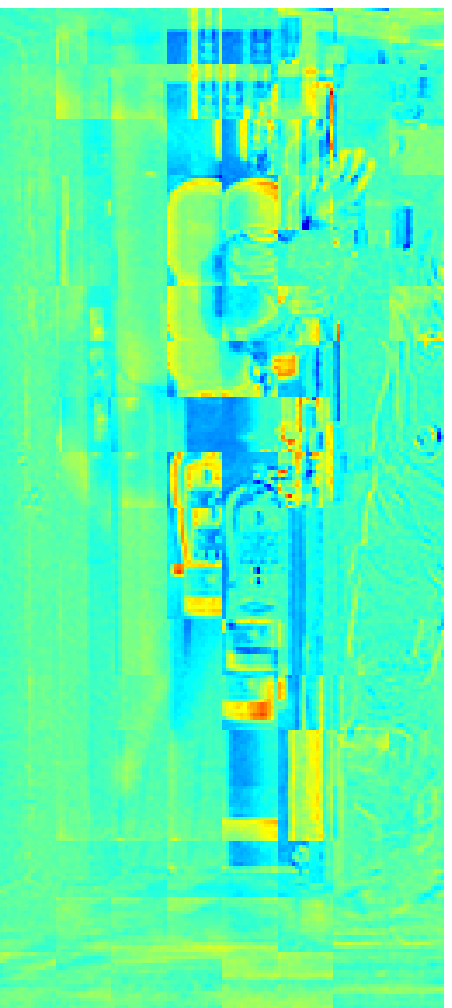}\hfill{}\includegraphics[width=0.11\textwidth]{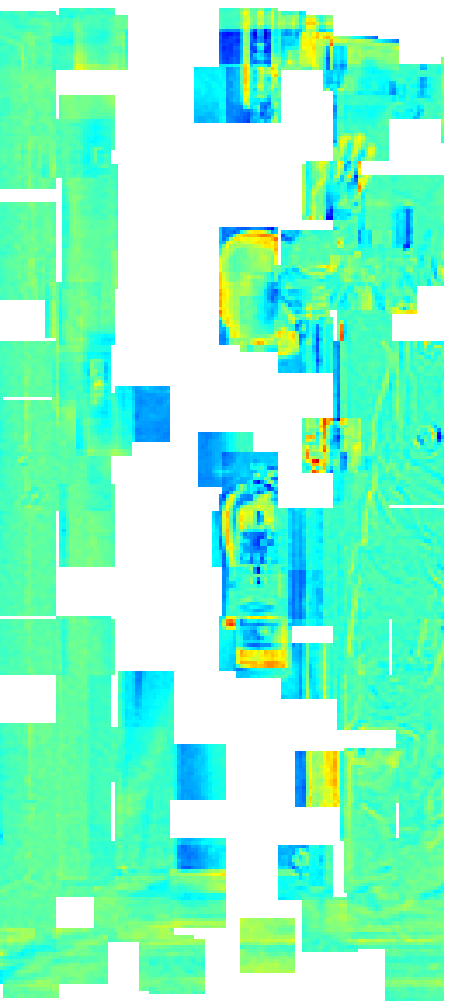}\hfill{}\includegraphics[width=0.11\textwidth]{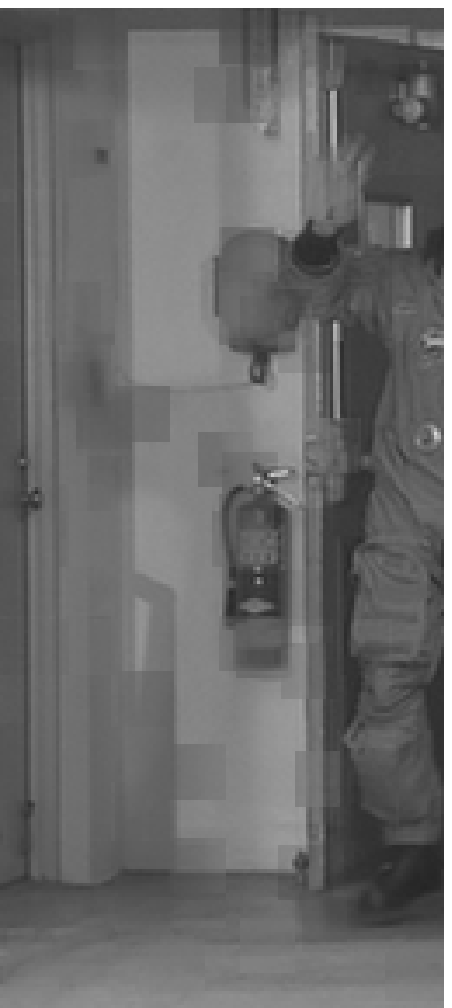}\hfill{}\includegraphics[width=0.11\textwidth]{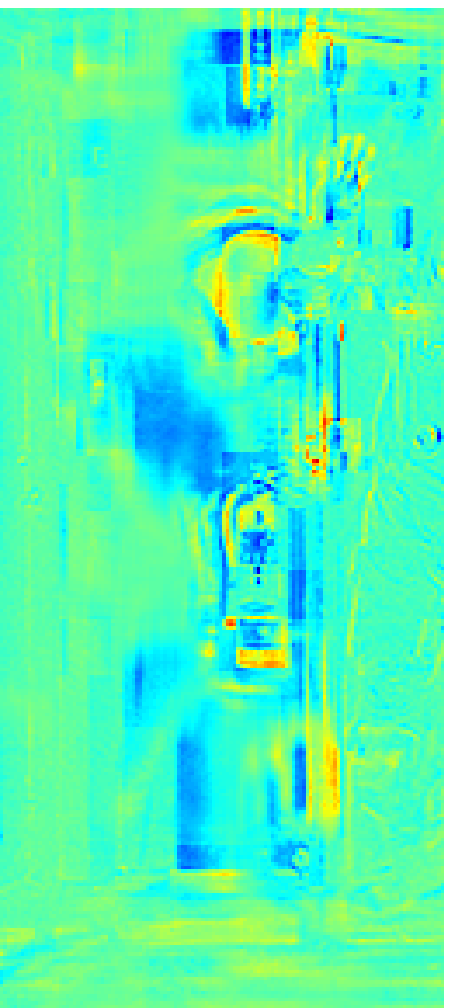}\hfill{}\includegraphics[width=0.11\textwidth]{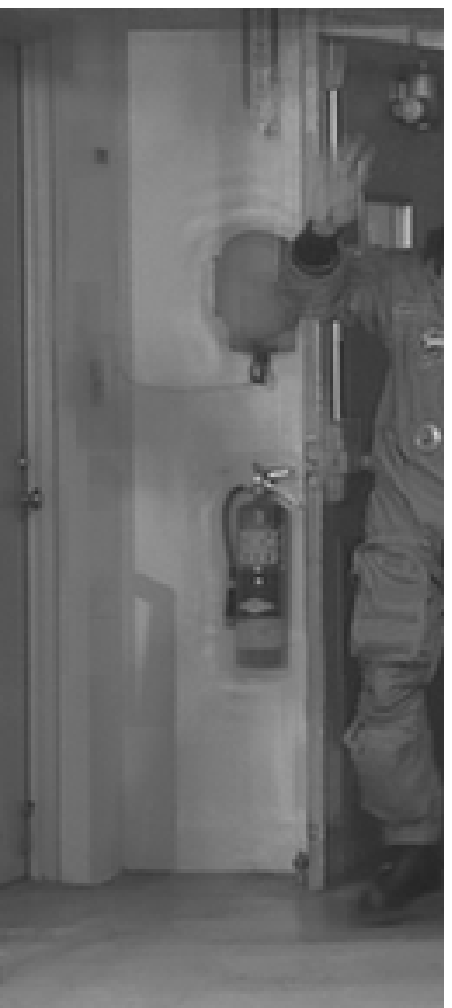}\vspace{-1mm}

\begin{minipage}[t]{0.11\textwidth}%
\begin{center}
$\sigf_{2t-1}$
\par\end{center}%
\end{minipage}\hfill{}%
\begin{minipage}[t]{0.11\textwidth}%
\begin{center}
$\sigf_{2t}$
\par\end{center}%
\end{minipage}\hfill{}%
\begin{minipage}[t]{0.11\textwidth}%
\begin{center}
$\pred_{2t}$
\par\end{center}%
\end{minipage}\hfill{}%
\begin{minipage}[t]{0.11\textwidth}%
\begin{center}
$\HP_{t}$
\par\end{center}%
\end{minipage}\hfill{}%
\begin{minipage}[t]{0.11\textwidth}%
\begin{center}
$\upd_{2t}$
\par\end{center}%
\end{minipage}\hfill{}%
\begin{minipage}[t]{0.11\textwidth}%
\begin{center}
$\LP_{t,\text{block}}$
\par\end{center}%
\end{minipage}\hfill{}%
\begin{minipage}[t]{0.11\textwidth}%
\begin{center}
$\updSE_{2t}$
\par\end{center}%
\end{minipage}\hfill{}%
\begin{minipage}[t]{0.11\textwidth}%
\begin{center}
$\LP_{t,\text{block+FSE}}$
\par\end{center}%
\end{minipage}

\hfill{}(a)\hfill{}\hfill{}(b)\hfill{}\hfill{}(c)\hfill{}\hfill{}(d)\hfill{}\hfill{}(e)\hfill{}\hfill{}(f)\hfill{}\hfill{}(g)\hfill{}\hfill{}(h)\hfill{}

\psfragscanoff

\spaceBeforeLabel

\caption{\label{fig:blockbased_inv}Details from the sequence \textit{crew}:
(a) reference frame, (b) current frame, (c) predictor, (d) compensated
highpass (heat map, green=0), (e) traditional update (heat map, green=0,
white:unconnected pixels), (f) traditionally resulting lowpass band,
the unconnected pixels in (e) lead to annoying artifacts in the lowpass
frame (f), (g) proposed reconstructed update (heat map, green=0),
(h) resulting lowpass band for proposed scheme}

\spaceBelowFig
\end{figure*}
compensation methods are used. There are other methods like mesh-based
approaches that show similar or even superior performance to block-based
methods but may have the drawback of a blurred predictor. In this
paper we concentrate on block-based compensation methods.

A block-based predictor $\pred_{2t}$ for the current frame $\sigf_{2t}$
is computed by searching for every block in the current frame for
a best fitting block in the reference frame $\sigf_{2t-1}$ in a specific
search window. As cost function we minimize the sum of squared differences
(SSD). 

The inversion of a block-based compensation leads to pixels that are
one-connected, multiple-connected or unconnected \cite{ohm1993,ohm2004interframe}.
The colored blocks in \fig{}~\ref{fig:Lifting-structure} are for
illustrating the occurrence of the three different cases of connectivity.
In the inverse compensated highpass band $\upd_{2t}$, the areas of
overlapped blocks show multiple-connected pixels while the white areas
show unconnected pixels. The remaining pixels are one-connected. Several
methods have been proposed for treating these cases.

In \cite{Choi1999}, the first candidate block is used for the update
of multiple-connected pixels while \cite{PPopescu2001} proposes to
use the candidate block with the smallest sum of absolute differences
which increases the similarity of the lowpass band and the corresponding
original frames. An optimum inversion regarding the reconstruction
error is proposed in \cite{girod2005} and \cite{tillier2005} by
analytically calculating the weights $a_{k}$ for multiple-connected
pixels.

For unconnected pixels, there is no information available. Up to now,
they are just copied from the reference frame without an update, as
proposed in \cite{ohm1993}. We observed that blocking artifacts occur
at the boundaries of the unconnected areas which makes a processing
of these areas necessary. In \cite{bozinovic2005}, interpolation
methods on the motion vector field are proposed to incorporate the
physical properties in addition to the reconstruction error.

For the inversion that is necessary for the update step, the blocks
are moved back to their original position. Thereby, we use the weights
from \cite{girod2005} for the one-connected and the multiple connected
pixels for averaging. In general, $k\text{-connected}$ pixels are
weighted by $a_{k}=\frac{1}{k+1}$ before they are added to the reference
frame.

\fig{}~\ref{fig:blockbased_inv} shows the intermediate results
for a detail from the sequence \textit{crew} for demonstrating the
reason of the occurring artifacts. \fig{}~\ref{fig:blockbased_inv}~(a)
shows the reference frame and \fig{}~\ref{fig:blockbased_inv}~(b)
shows the current frame. There is a change in the illumination between
these frames due to a flash light. The predictor that is computed
for the current frame is shown in \fig{}~\ref{fig:blockbased_inv}~(c).
The resulting highpass band is computed by applying \eqref{eq:H-Haar}
and is shown in \fig{}~\ref{fig:blockbased_inv}~(d). The inverse
compensated highpass frame for the update $\upd_{2t}$ according to
\eqref{eq:L-Haar} is shown in \fig{}~\ref{fig:blockbased_inv}~(e).
The white areas in the middle corresponds to unconnected pixels. This
update is added to the reference frame so the resulting lowpass frame,
shown in \fig{}~\ref{fig:blockbased_inv}~(f), shows block artifacts
in the areas of the unconnected pixels.

\section{Reconstruction of Unconnected Pixels}

The unconnected pixels can be regarded as holes in the update frame
$\upd_{2t}$ as shown by white areas in \fig{}~\ref{fig:blockbased_inv}~(e).
These pixels usually occur in areas of \disocclusion{} and occlusion.
It is hard to make any assumption about the actual motion in these
areas. So in opposite to the mentioned methods that try to interpolate
the motion vectors, we propose a different approach based on a signal
reconstruction method.

For this we use the Frequency Selective Extrapolation (FSE) \cite{seiler2010}
which can be used for reconstructing lost areas in image and video
data. It was shown in \cite{troeger2011}, that the FSE can also be
used for processing high frequency images. FSE is an iterative method
that generates a model 
\[
g\left[m,n\right]=\sum_{k\in\mathcal{K}}\hat{c}_{k}\varphi_{k}\left[m,n\right]
\]
for the unknown pixels based on the available pixels in $\upd_{2t}$.
For this, a weighted superposition of 2-D Fourier basis functions
$\varphi_{k}$ is generated where in every iteration the influence
$\hat{c}_{k}$ of the basis function that reduces the approximation
error the most is increased. For detailed description of FSE together
with pseudo code, please refer to \cite{seiler2010}.

The finally reconstructed pixels are illustrated by the dark areas
of $\updSE_{2t}$ in \fig{}~\ref{fig:Lifting-structure} and the
result of the reconstructed update frame $\updSE_{2t}$ is shown in
\fig{}~\ref{fig:blockbased_inv}~(g). The resulting lowpass frame
\begin{table*}[!t]
\begin{center}
%\begin{tabular*}{1\linewidth}{@{\extracolsep{\fill}}|c||c|c||c|c|c||c|c|}
\begin{tabular}{|c||>{\centering}p{1.3cm}|>{\centering}p{1.3cm}||>{\centering}p{1.3cm}|>{\centering}p{1.3cm}|>{\centering}p{1.3cm}||>{\centering}p{2cm}|>{\centering}p{2cm}|}
%>{\centering}p{1cm}|>{\centering}p{4cm}
\hline
 & \multicolumn{2}{c||}{

overall filesize in MB}& \multicolumn{3}{c||}{filesize for the lowpass
band only in MB}& \multicolumn{2}{c|}{mean PSNR in dB of the lowpass
band}\tabularnewline
\hline
& block & block+FSE & block & block+FSE & \% diff & block & block+FSE \tabularnewline

\hline
\hline

\textit{crew} &  \input{tab/YUV__3.1__crew__fs_complete.tex} & \input{tab/YUV__3.1__crew__fs_Lband.tex} & \input{tab/YUV__3.1__crew__psnr_Lband.tex}
\tabularnewline 

\hline

\textit{foreman} &  \input{tab/YUV__7.1__foreman__fs_complete.tex} & \input{tab/YUV__7.1__foreman__fs_Lband.tex} & \input{tab/YUV__7.1__foreman__psnr_Lband.tex}
\tabularnewline 

\hline

\textit{orient} &  \input{tab/YUV__14.1__discovery_orient__fs_complete.tex} & \input{tab/YUV__14.1__discovery_orient__fs_Lband.tex} & \input{tab/YUV__14.1__discovery_orient__psnr_Lband.tex}
\tabularnewline 

\hline

\textit{vimto} &  \input{tab/YUV__22.1__vimto__fs_complete.tex} & \input{tab/YUV__22.1__vimto__fs_Lband.tex} & \input{tab/YUV__22.1__vimto__psnr_Lband.tex}
\tabularnewline 

\hline

\textit{ClassA:People} & \input{tab/YUV__1000.1__ClassA-PeopleOnStreet__fs_complete.tex} & \input{tab/YUV__1000.1__ClassA-PeopleOnStreet__fs_Lband.tex} & \input{tab/YUV__1000.1__ClassA-PeopleOnStreet__psnr_Lband.tex}
\tabularnewline 

\hline

\textit{ClassA:Traffic} &  \input{tab/YUV__1001.1__ClassA-Traffic__fs_complete.tex} & \input{tab/YUV__1001.1__ClassA-Traffic__fs_Lband.tex} & \input{tab/YUV__1001.1__ClassA-Traffic__psnr_Lband.tex}
\tabularnewline 

\hline

\multicolumn{8}{c}{\vspace{-2.5mm}
}\tabularnewline 

\hline

\textit{cardiac} time &  \input{tab/MHA__1.130__heart_time__fs_complete.tex} & \input{tab/MHA__1.130__heart_time__fs_Lband.tex} & \input{tab/MHA__1.130__heart_time__psnr_Lband.tex}
\tabularnewline 

\hline

\textit{cardiac} slice &  \input{tab/MHA__2.10__heart_spat__fs_complete.tex} & \input{tab/MHA__2.10__heart_spat__fs_Lband.tex} & \input{tab/MHA__2.10__heart_spat__psnr_Lband.tex}
\tabularnewline 

\hline

\textit{thorax1} &  \input{tab/MHA__3.1__thorax1__fs_complete.tex} & \input{tab/MHA__3.1__thorax1__fs_Lband.tex} & \input{tab/MHA__3.1__thorax1__psnr_Lband.tex}
\tabularnewline 

\hline

\textit{thorax2} &  \input{tab/MHA__4.1__thorax2__fs_complete.tex} & \input{tab/MHA__4.1__thorax2__fs_Lband.tex} & \input{tab/MHA__4.1__thorax2__psnr_Lband.tex}
\tabularnewline 

\hline

\textit{head} &  \input{tab/MHA__5.1__head__fs_complete.tex} & \input{tab/MHA__5.1__head__fs_Lband.tex} & \input{tab/MHA__5.1__head__psnr_Lband.tex}
\tabularnewline 

\hline
\end{tabular}
\end{center}
\vspace{-3mm}

\caption{\spaceBeforeLabel\label{tab:tabular_results}Quality and overall
filesize for different methods, block and block+FSE include the rate
needed for the motion vectors}

\spaceBelowFig\spaceBelowFig
\end{table*}
of our proposed scheme is shown in \fig{}~\ref{fig:blockbased_inv}~(h)
where the block artifacts are suppressed. Hence, in our proposed scheme,
\eqref{eq:H-Haar} is left unchanged but \eqref{eq:L-Haar} is modified
to 
\begin{equation}
\LP_{t}=\sigf_{2t-1}+\left\lfloor \updSE_{2t}\right\rfloor \label{eq:L-Haar-1}
\end{equation}
where the reconstructed update $\updSE_{2t}$ is used for computing
the lowpass frame instead of the traditional update $\upd_{2t}$.

\section{\label{sec:Simulation-results}Simulation results}

We evaluated our proposed method with several video sequences, namely
\textit{crew}, \textit{foreman}, \textit{orient}, \textit{vimto},
and the HEVC test sequences \textit{ClassA:People} and \textit{ClassA:Traffic}.
The compensated transform can also be used for medical Computed Tomography
(CT) volumes \cite{schnurrer2012mmsp,schnurrer2012vcip} where adjacent
slices are taken as sequence. We evaluated our method using several
medical CT datasets, one \textit{head} and two \textit{thorax} 3-D
CT data sets%
\footnote{The CT volume data sets were kindly provided by Prof. Dr. med. Dr.
rer. nat. Reinhard Loose from the Klinikum Nürnberg Nord.%
} as well as a 4-D \textit{cardiac} volume%
\footnote{The CT volume data set was kindly provided by Siemens Healthcare.%
}. The slices of the CT data sets have a resolution of 512x512 with
32~slices (\textit{head}), 80~slices (\textit{thorax1}), 66~slices
(\textit{thorax2}), and 130~slices at 10 timesteps (\textit{cardiac}).
For the video sequences, we took the luminance component that has
a bit depth of 8~bit per pixel. The CT data sets have an intensity
component only that has a bit depth of 12~bit per voxel. The intensity
values describe the attenuation of the material at each voxel position. 

In our simulation, we perform one compensated Haar wavelet decomposition
step in slice respectively temporal direction and analyze the performance
of the proposed method for inverting the block-based compensation.
We use a blocksize of $16\times16$ and perform a full search for
each block within a search window of 15 pixels. For the FSE we use
the parameters according to \cite{seiler2010} except the maximum
number of iterations that was set to 1000. The resulting wavelet
coefficients are then coded frame by frame with JPEG~2000.

The first group of rows in \tab{}~\ref{tab:tabular_results} shows
the results for the video sequences and the second group of rows shows
the results for the CT data sets. The first two rows of the second
group show the results for the 4-D CT volume \textit{cardiac}, where
the first row shows the results for a transform in temporal direction
(time) while the second row shows the results for a transform in slice
direction (slice).

The first group of columns of \tab{}~\ref{tab:tabular_results}
headed by 'overall filesize' lists the filesize in MB needed for coding
the whole sequences listed in the first column in lossless mode. The
numbers in the columns headed by 'block' respectively 'block+FSE'
include the bits needed for motion vectors, as well. The reduction
in the filesize is obtained by avoiding the sharp edges of the block
artifacts. In the sequences \textit{foreman} and \textit{orient},
these sharp edges do not occur very often.

The second group of columns of \tab{}~\ref{tab:tabular_results}
headed by 'lowpass band only' lists the filesize for the lowpass image
in MB. The column '\%~diff' lists the relative difference between
the methods 'block' and 'block+FSE'. Negative values indicate that
'block+FSE' reduces the filesize. For most cases 'block+FSE' can reduce
the filesize compared to the traditional block-based method.

The third group of columns of \tab{}~\ref{tab:tabular_results}
headed by 'PSNR' lists the PSNR of the lowpass band compared to the
corresponding original frames for evaluating the quality. It is not
surprising that the obtained PSNR values are a little bit lower for
our proposed FSE-based method. Traditionally, the PSNR between the
frames in the lowpass band and the corresponding original frames is
one of the criteria for optimization. \fig{}~\ref{fig:Lifting-structure}
shows that the corresponding frame to the lowpass frame $\LP_{t}$
is the reference frame $\sigf_{2t-1}$. Assuming an optimum case with
a perfect prediction of the current frame, the highpass band \eqref{eq:H-Haar}
will become exactly zero. In this case, the lowpass band \eqref{eq:L-Haar}
will be identically the same as the reference slice $\sigf_{2t}$,
consequently resulting in an PSNR value of infinity. This is exactly
the case for unconnected areas as there is no update at all. So, traditionally
for unconnected pixels, no \emph{error} is added in the update step.
No matter how good the processing of unconnected pixels works, as
soon as we add an update for the unconnected pixels, the PSNR will
decrease because the metric aims at an update equal to zero. But this
is not desirable as the result is then a truncated wavelet transform,
i.e., without an update step, and the lowpass frame contains no information
about the current frame in these areas.

Nevertheless, we observed that artifacts can occur at the boundaries
of unconnected pixels as shown in \fig{}~\ref{fig:blockbased_inv}~(f).
As \fig{}~\ref{fig:blockbased_inv}~(h) shows, we can improve the
visual quality by suppressing these artifacts although the PSNR will
decrease.
\begin{figure*}
\hfill{}\textit{foreman}\hfill{}\hfill{}\textit{head}\hfill{}\hfill{}\textit{cardiac}\hfill{}

\vspace{-1mm}

\hfill{}%
\begin{minipage}[t]{0.16\textwidth}%
\begin{center}
reference
\par\end{center}%
\end{minipage}\hfill{}\hfill{}%
\begin{minipage}[t]{0.16\textwidth}%
\begin{center}
lowpass block
\par\end{center}%
\end{minipage}\hfill{}\hfill{}%
\begin{minipage}[t]{0.16\textwidth}%
\begin{center}
reference
\par\end{center}%
\end{minipage}\hfill{}\hfill{}%
\begin{minipage}[t]{0.16\textwidth}%
\begin{center}
lowpass block
\par\end{center}%
\end{minipage}\hfill{}\hfill{}%
\begin{minipage}[t]{0.16\textwidth}%
\begin{center}
reference
\par\end{center}%
\end{minipage}\hfill{}\hfill{}%
\begin{minipage}[t]{0.16\textwidth}%
\begin{center}
lowpass block
\par\end{center}%
\end{minipage}\hfill{}

\includegraphics[width=0.15\textwidth]{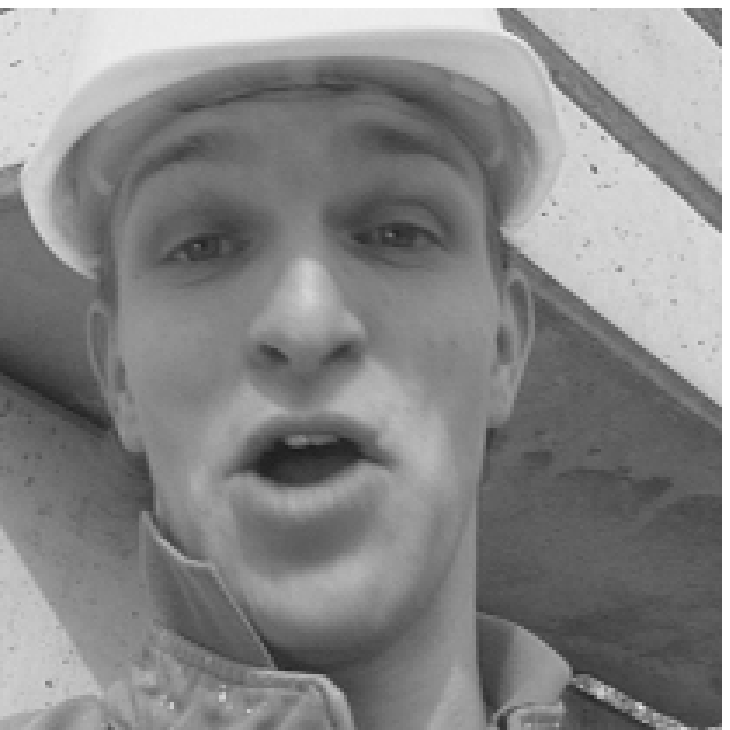}\hfill{}\includegraphics[width=0.15\textwidth]{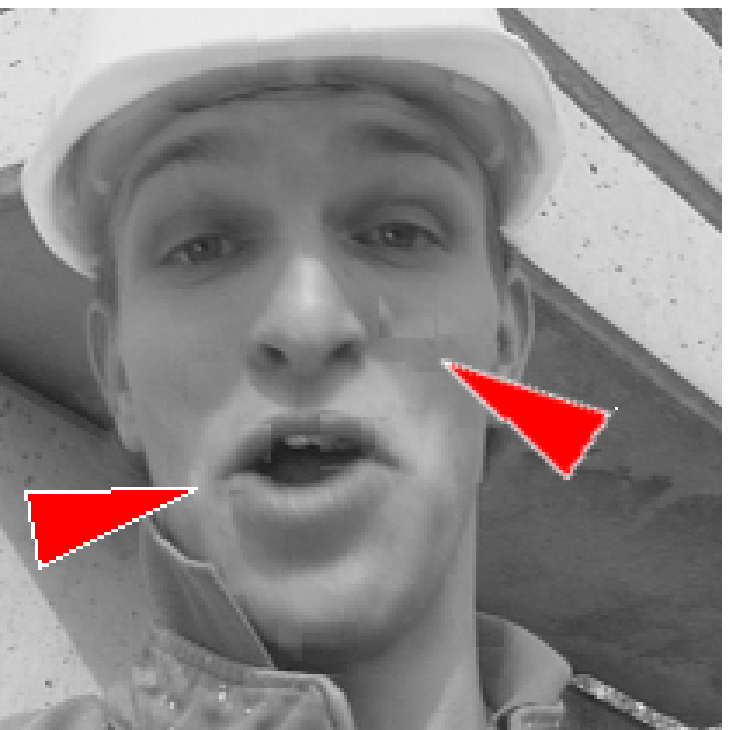}\hfill{}\includegraphics[width=0.15\textwidth]{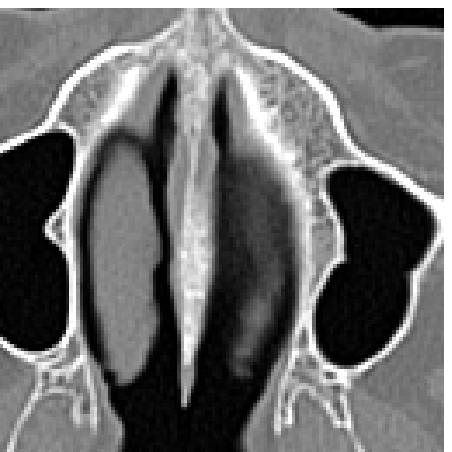}\hfill{}\includegraphics[width=0.15\textwidth]{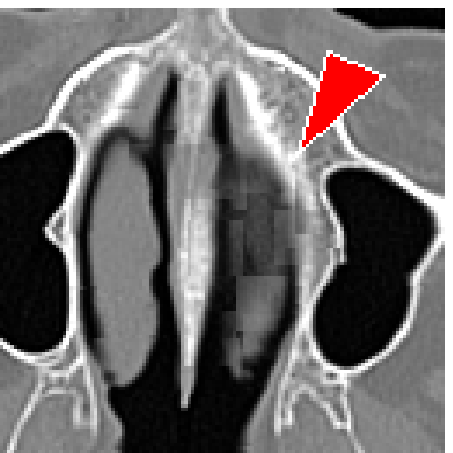}\hfill{}\includegraphics[width=0.15\textwidth]{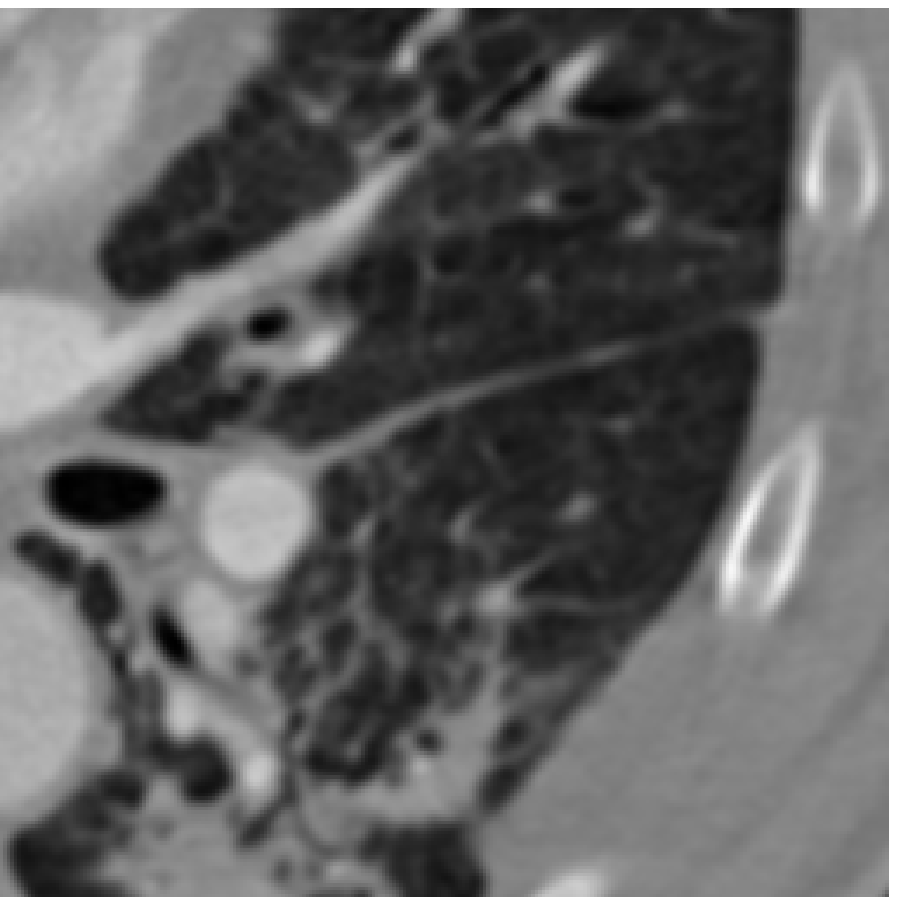}\hfill{}\includegraphics[width=0.15\textwidth]{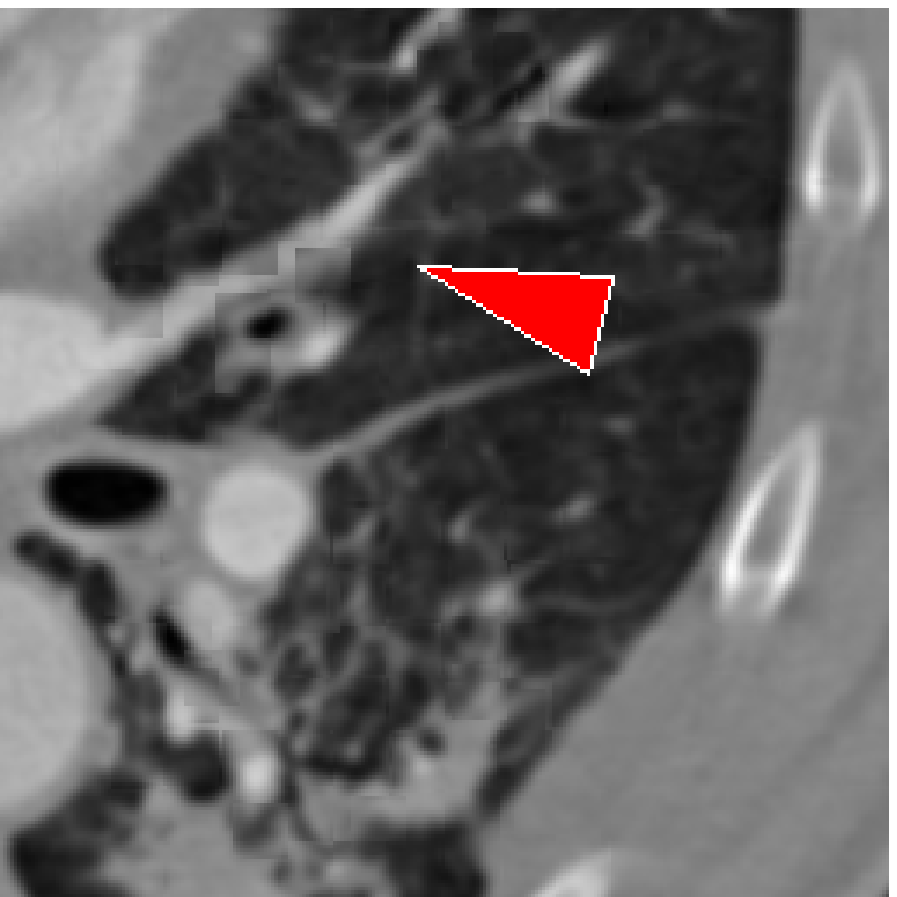}

\vspace{1mm}

\includegraphics[width=0.15\textwidth]{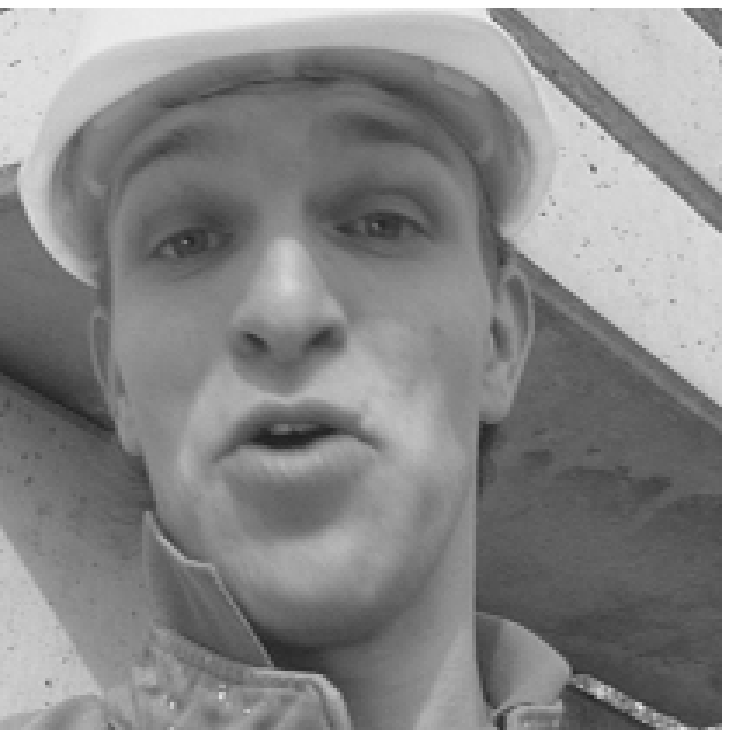}\hfill{}\includegraphics[width=0.15\textwidth]{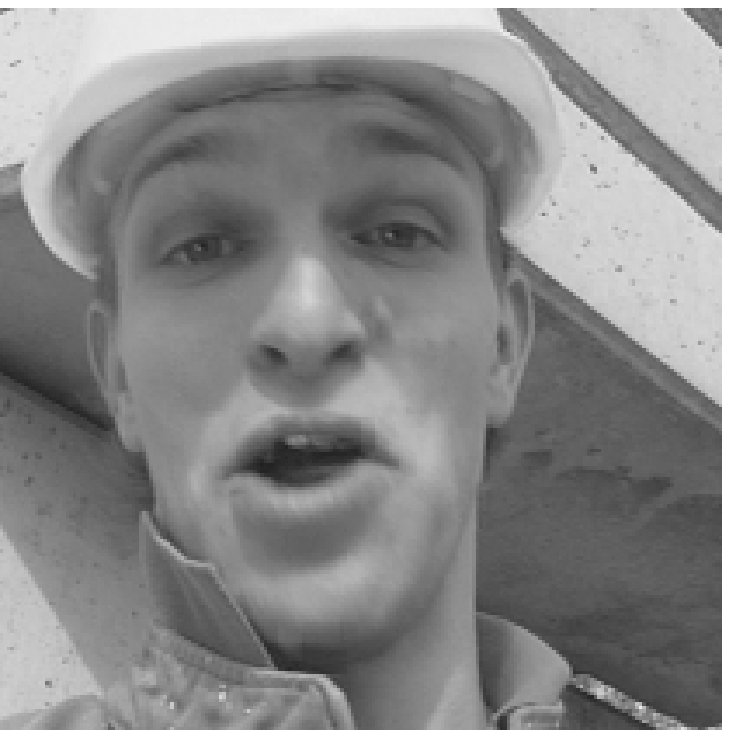}\hfill{}\includegraphics[width=0.15\textwidth]{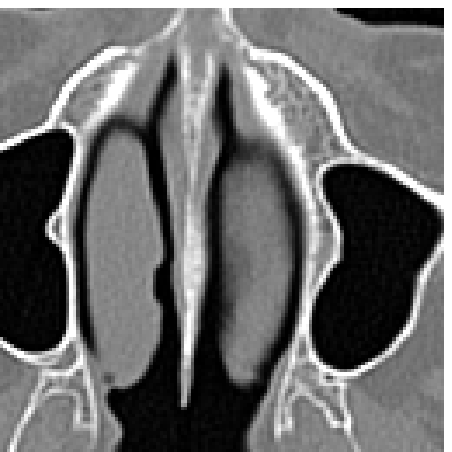}\hfill{}\includegraphics[width=0.15\textwidth]{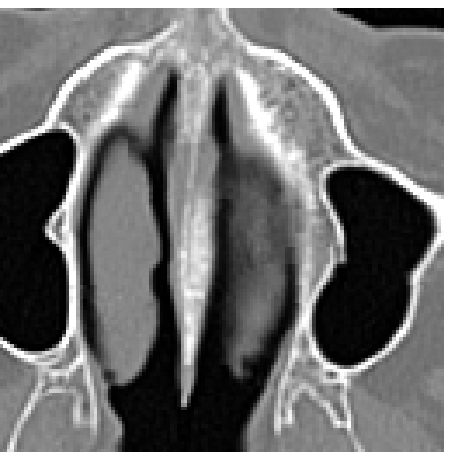}\hfill{}\includegraphics[width=0.15\textwidth]{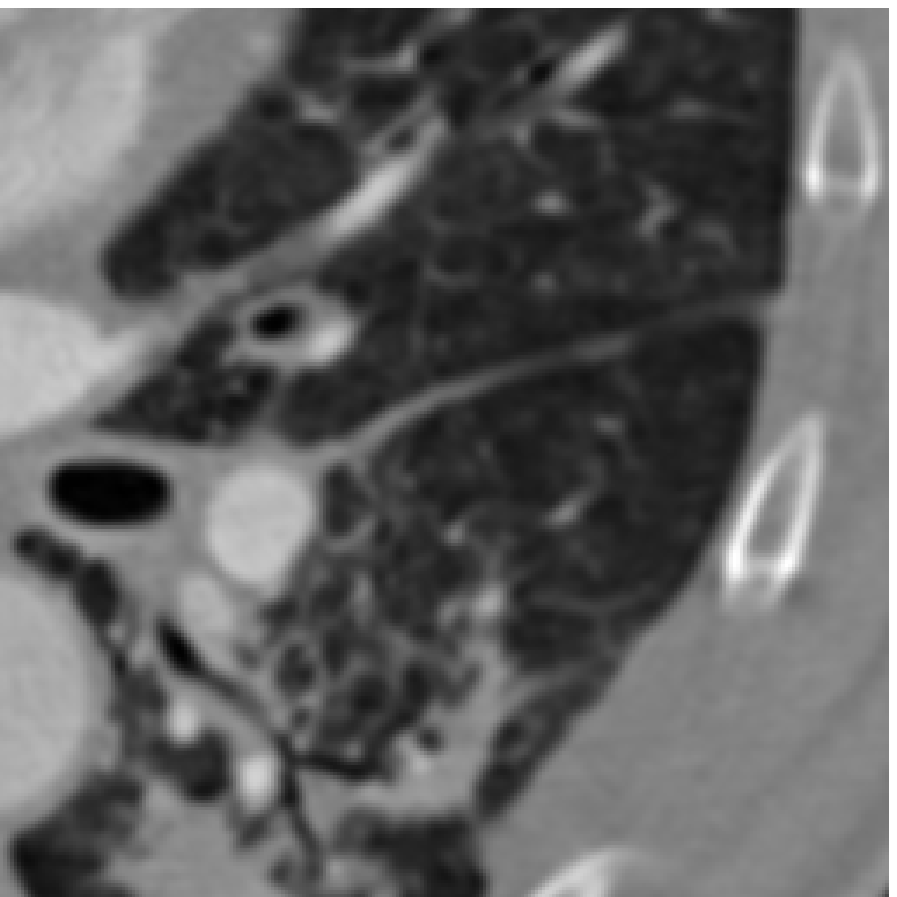}\hfill{}\includegraphics[width=0.15\textwidth]{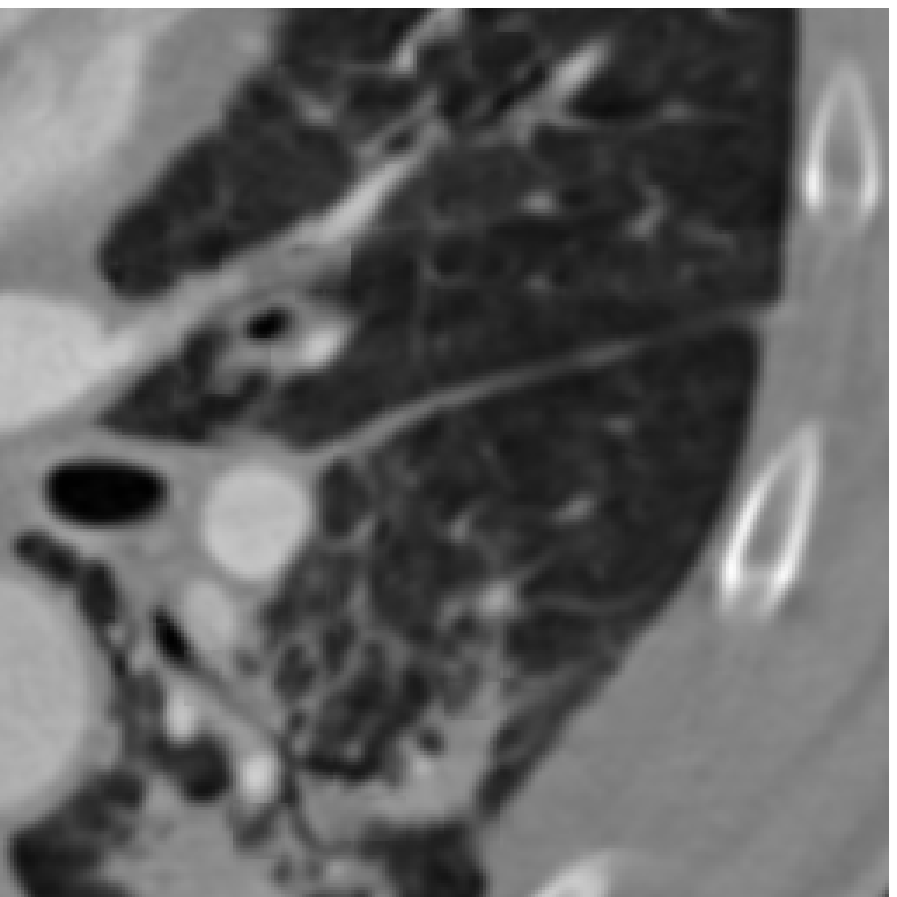}

\vspace{-1mm}
\hfill{}%
\begin{minipage}[t]{0.16\textwidth}%
\begin{center}
current
\par\end{center}%
\end{minipage}\hfill{}\hfill{}%
\begin{minipage}[t]{0.16\textwidth}%
\begin{center}
lowpass block+FSE
\par\end{center}%
\end{minipage}\hfill{}\hfill{}%
\begin{minipage}[t]{0.16\textwidth}%
\begin{center}
current
\par\end{center}%
\end{minipage}\hfill{}\hfill{}%
\begin{minipage}[t]{0.16\textwidth}%
\begin{center}
lowpass block+FSE
\par\end{center}%
\end{minipage}\hfill{}\hfill{}%
\begin{minipage}[t]{0.16\textwidth}%
\begin{center}
current
\par\end{center}%
\end{minipage}\hfill{}\hfill{}%
\begin{minipage}[t]{0.16\textwidth}%
\begin{center}
lowpass block+FSE
\par\end{center}%
\end{minipage}

\spaceBeforeLabel

\caption{\label{fig:examples}Details from the sequences \textit{foreman},
\textit{head} and \textit{cardiac}: block artifacts (marked by
red arrows) can be suppressed by the proposed scheme (block+FSE)}

\spaceBelowFig\vspace{-1mm}
\end{figure*}
 A perfect metric should evaluate how well the lowpass frame represents
all the frames in the reach of the wavelet filter and not how well
it fits to only one of them.

\fig{}~\ref{fig:examples} shows more examples of occurring block
artifacts (marked by red arrows) in the images denoted by 'lowpass
block' as well as the results of the reconstructed update frame denoted
by 'lowpass block+FSE'.

Further we can reduce the filesize for the lossless case compared
to the traditional compensated transform by avoiding the sharp edges
occurring at the boundaries of unconnected pixels. However, for the
medical CT volumes, the filesize increases when a compensation method
is used. This is caused by the correlated noise contained in these
data sets. Without a compensation method, the wavelet transform processes
adjacent pixels together. A compensated transform is applied according
to the structural information.

\section{Conclusion}

To obtain a high quality lowpass band without ghosting artifacts,
compensation methods have to be incorporated into the wavelet transform.
The block-based compensation method is a feasible compensation method
but the unconnected pixels have to be updated as well. Otherwise annoying
block artifacts can occur in the lowpass band and render it unusable
as downscaled version of the original sequence.

We showed that the Frequency Selective Extrapolation can be used for
creating an appropriate update for the unconnected pixels and improves
the visual quality of the lowpass band considerably. By avoiding block
artifacts in the lowpass band we can further reduce the filesize for
the lossless case compared to the traditional compensated transform
for video sequences and medical CT volumes.

Further work aims at the development of an appropriate metric for
evaluating the quality of the lowpass band.

\section*{Acknowledgment}

\small{We gratefully acknowledge that this work has been supported
by the Deutsche Forschungsgemeinschaft (DFG) under contract number
KA~926/4-1.}

\bibliographystyle{IEEEtran}
\bibliography{bibliography}


\end{document}

%% file: tab/YUV__3.1__crew__fs_complete.tex
26.35 & \textbf{26.29}

%% file: tab/YUV__3.1__crew__fs_Lband.tex
13.13 & \textbf{13.07} & -0.43

%% file: tab/YUV__3.1__crew__psnr_Lband.tex
\textbf{38.6} & 38.5

%% file: tab/YUV__7.1__foreman__fs_complete.tex
\textbf{20.89} & \textbf{20.89}

%% file: tab/YUV__7.1__foreman__fs_Lband.tex
\textbf{10.66} & 10.67 & +0.05

%% file: tab/YUV__7.1__foreman__psnr_Lband.tex
\textbf{37.7} & \textbf{37.7}

%% file: tab/YUV__14.1__discovery_orient__fs_complete.tex
\textbf{22.61} & 22.62

%% file: tab/YUV__14.1__discovery_orient__fs_Lband.tex
\textbf{10.90} & 10.91 & +0.10

%% file: tab/YUV__14.1__discovery_orient__psnr_Lband.tex
\textbf{40.2} & 40.1

%% file: tab/YUV__22.1__vimto__fs_complete.tex
24.09 & \textbf{24.04}

%% file: tab/YUV__22.1__vimto__fs_Lband.tex
11.40 & \textbf{11.35} & -0.45

%% file: tab/YUV__22.1__vimto__psnr_Lband.tex
\textbf{36.3} & 36.1

%% file: tab/YUV__1000.1__ClassA-PeopleOnStreet__fs_complete.tex
260.21 & \textbf{259.91}

%% file: tab/YUV__1000.1__ClassA-PeopleOnStreet__fs_Lband.tex
126.84 & \textbf{126.55} & -0.23

%% file: tab/YUV__1000.1__ClassA-PeopleOnStreet__psnr_Lband.tex
\textbf{35.4} & 35.3

%% file: tab/YUV__1001.1__ClassA-Traffic__fs_complete.tex
237.30 & \textbf{237.22}

%% file: tab/YUV__1001.1__ClassA-Traffic__fs_Lband.tex
122.32 & \textbf{122.23} & -0.07

%% file: tab/YUV__1001.1__ClassA-Traffic__psnr_Lband.tex
\textbf{41.9} & 41.8

%% file: tab/MHA__1.130__heart_time__fs_complete.tex
194.66 & \textbf{192.24}

%% file: tab/MHA__1.130__heart_time__fs_Lband.tex
95.00 & \textbf{92.58} & -2.61

%% file: tab/MHA__1.130__heart_time__psnr_Lband.tex
\textbf{47.9} & 47.7

%% file: tab/MHA__2.10__heart_spat__fs_complete.tex
192.54 & \textbf{190.41}

%% file: tab/MHA__2.10__heart_spat__fs_Lband.tex
94.89 & \textbf{92.76} & -2.30

%% file: tab/MHA__2.10__heart_spat__psnr_Lband.tex
\textbf{46.8} & 46.6

%% file: tab/MHA__3.1__thorax1__fs_complete.tex
14.70 & \textbf{14.63}

%% file: tab/MHA__3.1__thorax1__fs_Lband.tex
6.97 & \textbf{6.91} & -0.94

%% file: tab/MHA__3.1__thorax1__psnr_Lband.tex
\textbf{44.5} & \textbf{44.5}

%% file: tab/MHA__4.1__thorax2__fs_complete.tex
11.72 & \textbf{11.70}

%% file: tab/MHA__4.1__thorax2__fs_Lband.tex
5.67 & \textbf{5.65} & -0.33

%% file: tab/MHA__4.1__thorax2__psnr_Lband.tex
\textbf{45.9} & 45.6

%% file: tab/MHA__5.1__head__fs_complete.tex
7.33 & \textbf{7.32}

%% file: tab/MHA__5.1__head__fs_Lband.tex
3.56 & \textbf{3.55} & -0.28

%% file: tab/MHA__5.1__head__psnr_Lband.tex
\textbf{38.1} & \textbf{38.1}